\documentclass[prd,floatfix,onecolumn,amsmath,amssymb,10pt]{revtex4-2}
\usepackage{graphicx}
\usepackage[colorlinks=true,linkcolor=blue,citecolor=blue,urlcolor=blue]{hyperref}
\usepackage{float}    
\usepackage{placeins} 

\usepackage{caption} 
\usepackage[margin=1in]{geometry}
\usepackage[utf8]{inputenc}
\usepackage{cleveref}
\usepackage{amsmath,amssymb,bm}
\usepackage{graphicx,wrapfig,float,slashed,cancel}
\usepackage{subcaption}
\usepackage{placeins}
\usepackage{xcolor}
\usepackage{braket}
\usepackage{stmaryrd}
\usepackage{bbold} 
\usepackage{enumitem}
\usepackage{multirow}
\usepackage{orcidlink}
\usepackage{booktabs} 
\usepackage{epstopdf}

\definecolor{darkblue}{RGB}{1, 90, 173}

\begin{document}

\title{Study of $B \to K_0^*(1430)\,\ell^+ \ell^-$ decay in the standard model and scalar leptoquark scenario}

\author{M. Dadashzadeh\orcidlink{0009-0003-7008-2531}}
\author{K. Azizi\orcidlink{0000-0003-3741-2167}}
\email{kazem.azizi@ut.ac.ir}
\thanks{Corresponding author}

\affiliation{%
  \mbox{\textsuperscript{1}Department of Physics, University of Tehran, North Karegar Avenue, Tehran 14395--547, Iran}\\
  \mbox{\textsuperscript{2}Department of Physics,  Dogus University, Dudullu--Ümraniye, 34775 Istanbul, Türkiye}%
}

\date{\today}
\preprint{}
\begin{abstract}
This study examines the rare decay $B \to K_0^*(1430)\,\ell^+ \ell^-$ as a possible probe for new physics beyond the standard model (SM). We first analyze this channel within the SM and then include scalar leptoquark (LQ) contributions. We provide predictions for key observables, like differential decay rate, branching ratio, ratio of branching fractions at different channels, forward-backward asymmetry and different lepton polarizations,   and assess their sensitivity to leptoquark scenarios, highlighting $q^2$ regions less affected by the long-distance charmonium effects. The results can be useful  for future Belle II and LHCb measurements.
\end{abstract}
\maketitle

\renewcommand{\thefootnote}{\#\arabic{footnote}}
\setcounter{footnote}{0}

\section{Introduction}\label{intro}

The SM of particle physics provides a remarkably successful description of
elementary particles and their interactions over a wide range of energies. Its predictive
power has been confirmed in numerous colliders and low-energy experiments, resulting in
the discovery of the Higgs boson in 2012. Nevertheless, the SM is widely considered to be an
effective low-energy theory rather than a fundamental framework. It overlooks for the
existence of dark matter and dark energy, the origin of the baryon asymmetry, neutrino oscillations, and a unified description of quark
and lepton flavors. These theoretical and experimental shortcomings strongly motivate the
search for new physics (NP) beyond the SM. For general reviews of flavor physics and rare
$B$ decays, see, e.g., Refs.~\cite{Browder:2008em,Richman:2007zz,Owen:2014gda}.

Rare flavor-changing neutral current (FCNC) decays, in particular those mediated by the
quark-level transition $b \to s \ell^+\ell^-$, are among the most sensitive probes of NP.
In the SM,  such processes arise only at the loop level through penguin and box diagrams,
and their short-distance contributions are suppressed by the
Glashow-Iliopoulos-Maiani (GIM) mechanism~\cite{Buchalla:1995vs,Buras:1993xp,Buras:1994dj}.
Consequently, the corresponding branching ratios are small, and these decays are highly sensitive to contributions from heavy virtual particles. Theoretical descriptions are formulated in terms of the weak
effective Hamiltonian, where high-scale electroweak and possible NP dynamics are encoded in
Wilson coefficients $C_i(\mu)$, and operator mixing is governed by renormalization-group
evolution~\cite{Buras:1998AA,Isidori:2010KG,Hurth:2003VB}. See also a early phenomenological
analysis in Ref.~\cite{Ali:2002jg}.

Over the past decade, measurements of semileptonic $B$ decays have revealed discrepancies from SM predictions. The LHCb Collaboration has reported persistent deviations in
angular observables of the decay $B \to K^*(892)\mu^+\mu^-$, most prominently in the optimized
observable $P'_5$~\cite{LHCb:2013ghj,LHCb:2015svh,Bobeth:2007dw}, which have been tested with larger
datasets and also by other experiments~\cite{LHCb:2020lmf,LHCb:2022vje,CMS:2018qmi,ATLAS:2018cur}. Global
analyses of branching ratios and angular distributions reinforce these anomalies and suggest
NP contributions predominantly in the semileptonic Wilson coefficient $C_9$ (and possibly
$C_{10}$)~\cite{Altmannshofer:2014rta,Capdevila:2017bsm,Hiller:2003js,Descotes:2013wgg}.

Even more compelling are tests of lepton-flavor universality (LFU) in FCNC decays. Ratios such as
\begin{equation}
R_{K^{(*)}} =
\frac{\mathcal{B}(B\to K^{(*)}\mu^+\mu^-)}{\mathcal{B}(B\to K^{(*)}e^+e^-)},
\end{equation}
are theoretically very clean in the SM and are predicted to be extremely close to unity,
up to small electromagnetic corrections. However LHCb measurements of $R_K$ and $R_{K^*}$ have exhibited
deviations at the level of $2$--$3\sigma$~\cite{LHCb:2014vgu,LHCb:2017avl,Langenbruch:2018vuv}, generating considerable theoretical interest. More recent LHCb and CMS analyses based on larger datasets and improved
systematics have reduced some of these tensions~\cite{LopezHuertas:2024vif,CMS:2024syx}, yet
global fits to the full set of $b \to s\ell^+\ell^-$ data still favor NP scenarios that
modify the vector-current Wilson coefficients, especially $C_9$, in a lepton-flavor
non-universal way~\cite{Altmannshofer:2021qrr,Alguero:2019ptt,Khodjamirian:2012zq}. Besides, recent SMEFT global fits to $b\to s\ell^+\ell^-$ data, including both lepton
flavor universal and non-universal scenarios, are presented in
Ref.~\cite{Ali:2025xkw}.

Independent evidence for possible NP in the lepton sector arises from the long-standing
discrepancy between the measured and SM-predicted values of the muon anomalous magnetic
moment $a_\mu = (g_\mu-2)/2$. The Fermilab Muon $g-2$ experiment has validated and refined
the earlier Brookhaven result~\cite{Li:2021bbf,Muong-2:2025xyk}, thereby increasing the
tension with the state-of-the-art SM prediction based on high-precision hadronic inputs.
The muon $g-2$ anomaly, together with the hints of LFU violation in $B$ decays, motivates
NP scenarios that couple non-universally to leptons and can simultaneously affect low-energy
flavor observables and high-precision magnetic moment measurements.

For $b \to s\ell^+\ell^-$ transitions, an additional challenge is the control of
long-distance hadronic effects, in particular those induced by intermediate charmonium
resonances such as $J/\psi$ and $\psi'$ which strongly distort the dilepton invariant-mass
spectrum when $q^2\approx m_{c\bar c}^2$. These contributions are encoded in the charm-loop
piece of the effective coefficient $C_9^{\mathrm{eff}}$ and have been extensively studied
using QCD factorization, light-cone sum rules and dispersion relations
(e.g.~\cite{Khodjamirian:2010vf,Jager:2012uw,Lyon:2014hpa,Bobeth:2017vxj}). To reduce
the associated theoretical uncertainties, phenomenological analyses often focus on
short-distance $q^2$ windows that exclude the narrow charmonium peaks.

Most experimental and theoretical work to date has concentrated on pseudoscalar and vector
final states such as $B \to K\ell^+\ell^-$ and $B \to K^*(892)\ell^+\ell^-$ while more recent studies have also explored tensor
final states such as $B_s\to f_2'(1525)\mu^+\mu^-$~\cite{Rajeev:2020aut}. By contrast,
scalar final states like $K_0^*(1430)$ have received comparatively less attention, despite its complementary sensitivity to the effective Hamiltonian's chiral structure. 
The decay $B \to K_0^*(1430)\,\ell^+\ell^-$ proceeds via the same underlying
$b \to s \ell^+\ell^-$ transition but probes a different set of hadronic form factors and
operator structures. In particular, scalar final states lead to a simpler angular
distribution and can enhance the impact of scalar and pseudoscalar operators that are
either absent or helicity suppressed in vector modes
(see, e.g.,~\cite{Aslam:2009cv,Wang:2008da,Chen:2005an}). This makes
$B \to K_0^*(1430)\ell^+\ell^-$ an especially promising channel to explore NP with scalar
interactions and non-standard chiral couplings.

Within the broad classes of NP models proposed to explain the flavor anomalies, leptoquarks
(LQs) are particularly attractive. These hypothetical bosons carry both lepton and baryon
quantum numbers and appear naturally in a variety of ultraviolet completions such as
Pati-Salam unification, grand unified theories and models with composite fermions.
They induce tree-level contributions to semileptonic four-fermion operators and are
systematically classified according to their representations under
$SU(3)_C\times SU(2)_L \times U(1)_Y$~\cite{Buchmuller:1986zs}. Only a subset of the ten
possible scalar LQ multiplets is compatible with baryon-number conservation at the
renormalizable level. In the context of $b\to s\ell^+\ell^-$ anomalies, particular attention
has been paid to the two $SU(2)_L$ doublets
\[
X_{7/6} \equiv (3,2,7/6), \qquad X_{1/6} \equiv (3,2,1/6),
\]
which generate shifts in $C_9$, $C_{10}$ and their chirality-flipped
counterparts and can naturally produce LFU-violating patterns in $B$-meson decays
while remaining compatible with collider searches and other flavor constraints
(see, e.g.,~\cite{Dorsner:2016wpm,Arnold:2013cva,Davidson:1993qk,Fajfer:2012jt,Gripaios:2014tna,Crivellin:2017zlb}).

The study of $B \to K_0^*(1430)\ell^+\ell^-$ therefore offers a sensitive and complementary probe of the chiral and scalar structure of the $b\to s\ell^+\ell^-$ effective
Hamiltonian. In the SM, the forward-backward asymmetry in this decay vanishes identically
because of the absence of scalar couplings in the lepton sector~\cite{Aslam:2009cv};
any nonzero measurement would thus constitute a clean signal of NP. In addition,
lepton-polarization observables and LFU ratios in this channel probe in detail the
pattern of LQ-induced modifications to semileptonic operators and provide information
that is orthogonal to that from $B \to K^{(*)}\ell^+\ell^-$.

In this study we present a detailed analysis of the decay
$B \to K_0^*(1430)\,\ell^+\ell^-$ for $\ell=e,\mu,\tau$ within the SM and in a scalar
leptoquark scenario. We employ QCD sum-rule form factors for the
$B\to K_0^*(1430)$ transition~\cite{Aliev:2007rq} and implement both the short-distance SM
and NP contributions using the full effective Hamiltonian including chirality-flipped
operators. We compute differential decay rates, branching fractions, LFU ratios,
forward-backward asymmetries and lepton-polarization asymmetries, paying particular
attention to short-distance $q^2$ regions where long-distance charmonium effects are
minimized. Our numerical analysis shows that realistic scalar LQ benchmarks can induce
non-negligible and potentially observable deviations from SM expectations in several of these
observables. This highlights $B \to K_0^*(1430)\,\ell^+\ell^-$ as a promising mode for
future precision studies at Belle~II and the upgraded LHCb experiment, where the
non-resonant dilepton mass regions considered here will be experimentally accessible
with high statistics~\cite{BelleII:2019zhz,BelleII:2021tzi,Bediaga:2018roe}.

Before proceeding, we briefly outline the structure of the paper.
In Sec.~\ref{sec:Heff}, we review the effective Hamiltonian for the $b\to s\ell^+\ell^-$ transition in the SM and show how it is modified in the scalar-leptoquark scenario, including the chirality-flipped operators. In Sec.~\ref{ffs}, we introduce the $B\to K_0^*(1430)$ form factors and build the decay amplitude in terms of the form factors and Wilson coefficients. In Sec.~\ref{ddr}, we present the differential decay rate and illustrate the $q^2$ spectra for $\ell=e,\mu,\tau$ in the SM and with leptoquark contributions. In Sec.~\ref{sec:results_LFU}, we give numerical results for the branching fractions integrated over short-distance $q^2$ windows away from the $J/\psi$ and $\psi'$ resonances, and we discuss the LFU ratio $R_{K_0^*}$. In Sec.~\ref{fba},  we analyze angular and polarization observables, emphasizing the forward-backward asymmetry as a clean SM null test and showing the impact of leptoquarks on the longitudinal lepton polarization. Our conclusions are collected in Sec.~\ref{conclu}.

\section{Effective Hamiltonian}\label{sec:Heff}

Rare $b \to s\ell^+\ell^-$ transitions are most conveniently described within the framework
of an effective field theory obtained after integrating out the heavy degrees of freedom
(top quark, $W^\pm$, $Z$, and any heavy NP states). At scales $\mu \sim m_b$ the resulting
effective weak Hamiltonian can be written as
\begin{equation}
\mathcal{H}_{\text{eff}} = -4 \frac{G_F}{\sqrt{2}} V_{ts}^* V_{tb}
\sum_{i=1}^{10} C_i(\mu)\, O_i(\mu)\,.
\end{equation}
Here $G_F$ is the Fermi constant, $V_{ts}^*V_{tb}$ denotes the relevant combination of CKM
matrix elements, $C_i(\mu)$ are the Wilson coefficients, and $O_i(\mu)$ are local
operators built from quark and lepton fields. The Wilson coefficients encode the
short-distance physics and are evaluated in a given renormalization scheme (typically
naive Dimensional Regularization,  NDR) and evolved from the electroweak scale down to
$\mu\sim m_b$ by renormalization-group methods~\cite{Buras:1993xp,Buras:1994dj}.

Although the full basis contains current-current, QCD penguin, electroweak penguin,
and semileptonic operators, the dominant short-distance contributions to
$b\to s\ell^+\ell^-$ observables arise from the magnetic dipole and semileptonic operators
$O_7$, $O_9$, and $O_{10}$. The four-quark operators $O_{1\text{--}6}$ enter indirectly
through operator mixing and via long-distance $c\bar{c}$ contributions to the semileptonic
amplitude. Explicitly,
\begin{align}
O_7   &= \frac{e}{16\pi^2} m_b \left(\bar{s}_\alpha \sigma^{\mu\nu} R\, b_\alpha\right) F_{\mu\nu}, \\
O_9   &= \frac{e^2}{16\pi^2} \left(\bar{s}_\alpha \gamma^\mu L\, b_\alpha\right) \left(\bar{\ell}\gamma_\mu \ell\right), \\
O_{10} &= \frac{e^2}{16\pi^2} \left(\bar{s}_\alpha \gamma^\mu L\, b_\alpha\right) \left(\bar{\ell}\gamma_\mu \gamma_5 \ell\right),
\end{align}
where $L=(1-\gamma_5)/2$ and $R=(1+\gamma_5)/2$ denote the left and right-handed
chirality projectors. The operator $O_7$ drives radiative decays such as
$B\to K^*\gamma$ and contributes to $B\to K^{(*)}\ell^+\ell^-$ through a virtual photon,
while $O_{9,10}$ represent vector and axial-vector semileptonic interactions.

In terms of these operators, the effective Hamiltonian relevant for
$b \to s \ell^+\ell^-$ can be written as
\begin{equation}
\begin{aligned}
{\cal H}_{\text{eff}}
= \frac{G_F \alpha_{\text{em}} V_{tb} V_{ts}^\ast}{2\sqrt{2} \pi}
\Bigg[
& C_9^{\text{eff}}(m_b)\, \bar{s}\gamma_\mu (1-\gamma_5) b \; \bar{\ell} \gamma^\mu \ell
+ C_{10}(m_b)\, \bar{s} \gamma_\mu (1-\gamma_5) b \; \bar{\ell} \gamma^\mu \gamma_5 \ell \\
& - \frac{2 m_b}{q^2}\, C^{eff}_7(m_b)\, \bar{s}\, i \sigma_{\mu\nu} q^\nu (1+\gamma_5) b \; \bar{\ell} \gamma^\mu \ell
\Bigg].
\end{aligned}
\end{equation}

The coefficient $C_9^{\text{eff}}$ contains both a perturbative short-distance part and
long-distance contributions from $c\bar c$ intermediate states. These long-distance effects
enter through the matrix elements of four-quark operators
$\langle \ell^+ \ell^- s \,|\, \mathcal{O}_i \,|\, b \rangle$ with $1 \leq i \leq 6$ and can be
absorbed into an effective coefficient
\begin{equation}
\label{eq:C9eff}
C_{9}^{\text{eff}}(\hat{s}) = C_{9} + Y(\hat{s})\,, \qquad \hat s \equiv \frac{q^2}{m_B^2}.
\end{equation}
The Wilson coefficient $C_9$ is defined at the electroweak scale and evolved down to
$\mu\sim m_b$ via the renormalization group, while the function $Y(\hat s)$ encodes the
effect of four-quark operators when they couple to a virtual photon subsequently
producing the lepton pair~\cite{deBoer:2016dcg}. The perturbative part of $Y(\hat s)$ is
a calculable loop function with GIM suppression, whereas the long-distance part associated
with narrow charmonium resonances dominates in the vicinity of $q^2 \simeq m_{J/\psi}^2$
and $m_{\psi'}^2$ and cannot be reliably described by fixed-order perturbation
theory~\cite{Faustov:2018dkn}. In practice, one often augments the perturbative charm-loop
contribution by phenomenological Breit-Wigner terms; here, however, we focus on the
short-distance regions and explicitly exclude the narrow-resonance windows.

Beyond the SM, new heavy states can generate additional operators or modify the Wilson
coefficients of the existing ones. In particular, scalar leptoquarks can induce
chirality-flipped operators $O_{7,9,10}'$ with opposite quark chiralities. A convenient
generalized Hamiltonian including these structures is
\begin{equation}
\label{eq:SM_eff_hamiltonian}
\begin{aligned}
\mathcal{H}_{\text{eff}} =
\frac{G_{F}\alpha_{\text{em}}V_{tb}V^{*}_{ts}}{2\sqrt{2}\pi}
\Big[
& C_{9}^{\text{eff}}\, \bar{s}\gamma_{\mu}(1-\gamma_{5})b\, \bar{\ell}\gamma^{\mu}\ell
+ C_{9}^{\prime\,\text{eff}}\, \bar{s}\gamma_{\mu}(1+\gamma_{5})b\, \bar{\ell}\gamma^{\mu}\ell \\
&+ C_{10}\, \bar{s}\gamma_{\mu}(1-\gamma_{5})b\, \bar{\ell}\gamma^{\mu}\gamma_{5}\ell
+ C_{10}^{\prime}\, \bar{s}\gamma_{\mu}(1+\gamma_{5})b\, \bar{\ell}\gamma^{\mu}\gamma_{5}\ell \\
&- \frac{2m_{b}}{q^{2}}\, C_{7}^{\text{eff}}\, \bar{s}\, i \sigma_{\mu\nu} q^{\nu}(1+\gamma_{5}) b\, \bar{\ell}\gamma^{\mu}\ell
- \frac{2m_{b}}{q^{2}}\, C_{7}^{\prime\,\text{eff}}\, \bar{s}\, i \sigma_{\mu\nu} q^{\nu}(1-\gamma_{5}) b\, \bar{\ell}\gamma^{\mu}\ell
\Big].
\end{aligned}
\end{equation}
In the SM, the primed coefficients are highly suppressed and can be neglected to an
excellent approximation. NP can, however, introduce these chirality-flipped operators
together with their associated Wilson coefficients.

In the scalar LQ model under consideration, the Wilson coefficients are modified as
\cite{Arnold:2013cva,Azizi:2016dcj}:
\begin{equation}
C_{9}^{\text{LQ}} = C_{10}^{\text{LQ}} = - \frac{\pi}{2\sqrt{2}\, G_{F} \alpha_{em} V_{tb} V_{ts}^{*}}
\frac{\lambda_{e}^{23} \lambda_{e}^{22*}}{M_{Y}^{2}} ,
\end{equation}
\begin{equation}
C_{9}^{\prime\, \text{LQ}} = - C_{10}^{\prime\, \text{LQ}} =
\frac{\pi}{2\sqrt{2}\, G_{F} \alpha_{em} V_{tb} V_{ts}^{*}}
\frac{\lambda_{s}^{22} \lambda_{b}^{32*}}{M_{V}^{2}} .
\end{equation}
Here $\lambda_{e,s,b}^{ij}$ denote the relevant LQ Yukawa couplings and $M_{Y,V}$
their masses. The total effective coefficients appearing in the decay amplitude then read
\begin{align}
C_{9}^{\text{eff,tot}} &= C_{9}^{\text{eff}} + C_{9}^{\text{LQ}}, &
C_{10}^{\text{tot}} &= C_{10} + C_{10}^{\text{LQ}}, \\
C_{9}^{\prime\,\text{eff,tot}} &= C_{9}^{\prime\,\text{eff}} + C_{9}^{\prime\,\text{LQ}}, &
C_{10}^{\prime\,\text{tot}} &= C_{10}^{\prime} + C_{10}^{\prime\,\text{LQ}}.
\end{align}
For illustration, we quote representative benchmark ranges used later:
\begin{align}
C_{7}^{\text{eff}} &= -0.295, \\
C_{9}^{\text{eff}} &\in [1.573,\,6.625], \\
C_{10} &= -4.260, \\
C_{9}^{\text{eff,tot}} &\in [2.793,\,4.394], \\
C_{9}^{\prime\,\text{eff,tot}} &\in [0,\,1.586], \\
C_{10}^{\text{tot}} &\in [-5.846,\,-4.260], \\
C_{10}^{\prime\,\text{tot}} &\in [-1.586,\,0],
\end{align}
where $C_{9,10}^{\text{LQ}}$ receive contributions from $X^{(7/6)}=(3,2,7/6)$ and the primed
coefficients from $X^{(1/6)}=(3,2,1/6)$. Exact values depend on the LQ couplings and
masses; here they represent benchmark intervals consistent with current constraints.

Note the present work employs the standard phenomenological expression for the effective
coefficient $C_9^{\rm eff}(\hat{s})$ given in Eq.~(\ref{eq:C9eff}), which incorporates
the factorizable quark-loop contribution of the four-quark operators at leading order in
QCD. A fully systematic treatment within QCD factorization (QCDF) would additionally
require incorporating hard spectator-scattering and weak-annihilation contributions.
Regarding hard spectator scattering, the QCDF framework developed in
Ref.~\cite{Beneke:2001at} has been worked out explicitly for pseudoscalar and vector
final states such as $K$ and $K^*$, where the decay amplitudes receive hard spectator
contributions expressed as convolutions of perturbative kernels with the light-meson and
$B$-meson light-cone distribution amplitudes (LCDAs). For the scalar final state
$K_0^*(1430)$ studied here, an analogous QCDF analysis of hard spectator scattering has,
to the best of our knowledge, not yet been performed, owing to the different LCDA
structure of scalar mesons and the corresponding absence of a factorization theorem for
$B\to K_0^*(1430)\ell^+\ell^-$ that includes hard spectator scattering. Extending the
 formalism~\cite{Beneke:2001at} to scalar final states would require specifying the
leading-twist and twist-3 scalar-meson LCDAs and deriving the appropriate
hard-scattering kernels, which goes beyond the scope of the present paper.

Regarding weak annihilation, Huang \textit{et al.} in Ref.~\cite{Huang:2024xii} performed
the first NLO computation of the weak-annihilation contribution in a soft-collinear
effective theory (SCET) framework for channels with pseudoscalar and vector final states.
For the $b\to d$ electroweak-penguin decays considered there, the newly computed NLO
weak-annihilation contribution can bring about an $\mathcal{O}(35\%)$ reduction of the
real part and an $\mathcal{O}(15\%)$ reduction of the imaginary part of the effective
coefficient $C^{(u)}_{9,P}$ in the kinematic range $1.5\,\mathrm{GeV}^2\leq q^2\leq
4.0\,\mathrm{GeV}^2$. For the $b\to s$ transitions relevant to our analysis, the
corresponding weak-annihilation amplitudes are proportional to
$\lambda_u^{(s)}=V_{ub}V_{us}^*$, which is doubly Cabibbo-suppressed relative to
$\lambda_c^{(s)}=V_{cb}V_{cs}^*$; such effects may therefore be numerically smaller and
should be regarded as an additional, presently unquantified source of systematic
uncertainty. Furthermore, a dedicated weak-annihilation calculation for the scalar final
state $B\to K_0^*(1430)\ell^+\ell^-$ has not been carried out in the literature. The
existing NLO analyses are formulated for pseudoscalar and vector mesons, where the
helicity structure of the amplitude and the relevant LCDAs are well established. For
scalar mesons the situation differs: the vector-current decay constant $f_S$ is
suppressed and vanishes in the SU(3) limit~\cite{Wang:2008da}, so the leading-twist LCDA
is normalized through the scale-dependent scalar decay constant $\bar{f}_S(\mu)$, which
is significantly less constrained than its pseudoscalar or vector counterparts. In the
case of the strange scalar $K_0^*(1430)$ with $m_s\gg m_{u,d}$, the suppression is
moderate. Implementing weak annihilation at the same level of rigor as in
Refs.~\cite{Beneke:2001at,Huang:2024xii} for a scalar ($0^{++}$) final state would require
a dedicated extension of the existing QCDF/SCET frameworks to scalar final states,
together with a corresponding factorization formula involving scalar-meson LCDAs. To the
best of our knowledge, such a framework has not yet been developed in the literature.
Consequently, we do not attempt to model weak annihilation for $B\to
K_0^*(1430)\ell^+\ell^-$ in this work and regard it as a subleading but presently
unquantified source of systematic uncertainty, in addition to the dominant form-factor
uncertainties. A full QCDF analysis of $B\to K_0^*(1430)\ell^+\ell^-$ including hard
spectator scattering and weak annihilation constitutes an interesting direction for
future work.

\section{Form factors and decay amplitude}\label{ffs}

The semileptonic $B\to K_0^*(1430)$ transition form factors have been computed using
several methods. In particular, Ref.~\cite{Wang:2008da} provides a light-cone sum rule
(LCSR) determination based on scalar-meson LCDAs, while Ref.~\cite{Aliev:2007rq}
employs three-point QCD sum rules, and Ref.~\cite{Chen:2007na} uses a covariant
light-front quark model. As stressed in Ref.~\cite{Wang:2008da}, their LCSR form
factors are numerically about twice as large as those obtained in the other two
approaches. For example, for $\bar{B}^0\to K_0^*(1430)$ one finds
$f_+(0)=0.97^{+0.20}_{-0.20}$ in LCSR, to be compared with $f_+(0)=0.52$ in the
light-front quark model and $f_+(0)=0.62\pm0.16$ in the three-point sum-rule
calculation. This reflects genuine differences in the underlying nonperturbative
inputs and modeling, in particular the treatment of the scalar-meson LCDAs and the
kinematic extrapolation.

In the present work we adopt the three-point QCD sum-rule form factors of
Ref.~\cite{Aliev:2007rq} as our primary hadronic input. Numerically, these are in good
agreement with the covariant light-front quark model and thus supported by an
independent method, while leading to more conservative branching-fraction
predictions than those obtained with the larger LCSR inputs.

The hadronic dynamics in the exclusive decay $B \to K_0^*(1430)\ell^+\ell^-$ is encoded in
$B\to K_0^*(1430)$ transition form factors, defined through the matrix elements
\begin{equation}
\label{fone}
\langle K_{0}^{*}(1430)(p') | \bar{s}\gamma_{\mu}\gamma_{5} b | B(p) \rangle
= f_{+}(q^{2})\, \mathcal{P}_{\mu} + f_{-}(q^{2})\, q_{\mu} ,
\end{equation}
\begin{equation}
\label{ftwo}
\langle K_{0}^{*}(1430)(p') | \bar{s}\, i \sigma_{\mu\nu} q^{\nu} \gamma_{5} b | B(p) \rangle
= \frac{f_{T}(q^{2})}{m_{B} + m_{K_{0}^{*}}}
\left[ \mathcal{P}_{\mu}\, q^{2} - \bigl(m_{B}^{2} - m_{K_{0}^{*}}^{2}\bigr)\, q_{\mu} \right] ,
\end{equation}
with $\mathcal{P}_{\mu}= (p+p')_{\mu}$ and $q_{\mu}= (p-p')_{\mu}$. These form factors are
nonperturbative inputs and, in this work, we take them from QCD sum-rule calculations based on
three-point correlation functions~\cite{Aliev:2007rq}.

Concretely, one considers the correlators
\begin{equation}
\Pi_{\mu}(p,p',q)= i^{2} \int \mathrm{d}^4x\,\mathrm{d}^4y\;
e^{i p'\!\cdot y - i p\!\cdot x}\;
\bra{0}\, T\!\left[J^{K_0^*}(y)\, J_{\mu}^{A}(0)\, J_{5}^{B}(x)\right] \ket{0},
\end{equation}
and
\begin{equation}
\Pi_{\mu\nu}(p,p',q)= i^{2} \int \mathrm{d}^4x\,\mathrm{d}^4y\;
e^{i p'\!\cdot y - i p\!\cdot x}\;
\bra{0}\, T\!\left[J^{K_0^*}(y)\, J_{\mu\nu}(0)\, J_{5}^{B}(x)\right] \ket{0},
\end{equation}
where $J^{K_0^*}=\bar d\, s$ and $J_{5}^{B}=\bar b\, i\gamma_{5} d$ are interpolating currents for
$K_0^*(1430)$ and $B$, while $J_{\mu}^{A}=\bar s\,\gamma_{\mu}\gamma_{5} b$ and
$J_{\mu\nu}=\bar s\, i\sigma_{\mu\nu} b$ denote the transition currents. Matching the hadronic
representation of these correlators to their operator-product expansion (OPE) and applying a double
Borel transformation in $p^{2}$ and $p'^{2}$ yields sum rules for $f_{+}(q^{2})$, $f_{-}(q^{2})$, and
$f_{T}(q^{2})$.
Finally,  combining the matrix elements in Eqs.~\eqref{fone} and~\eqref{ftwo} with the generalized
effective Hamiltonian~\eqref{eq:SM_eff_hamiltonian}, the $B \to K_0^*(1430)\ell^+\ell^-$ decay
amplitude can be written directly in terms of the form factors and the short-distance Wilson
coefficients. It is then convenient to reorganize the result into a helicity-decomposed form,
\begin{align}
\label{eq:24}
\mathcal M(B \to K_0^*\,\ell^+\ell^-) &=
\frac{G_F \alpha_{\rm em}}{2\sqrt{2}\pi} V_{tb}V_{ts}^*
\Big\{
\bar\ell \gamma^\mu \ell \, \big[ X_V (p+p')_\mu + Y_V q_\mu \big]
+ \bar\ell \gamma^\mu\gamma_5 \ell \, \big[ X_A (p+p')_\mu + Y_A q_\mu \big]
\Big\},
\end{align}
with the invariant amplitudes
\begin{align}
X_V &= \big(C_9'^{\text{eff}} - C_9^{\text{eff}}\big)\, f_+(q^2)
      - \frac{2 m_b}{m_B+m_{K_0^*}}\big(C_7^{\text{eff}} - C_7'^{\text{eff}}\big)\, f_T(q^2), \label{eq:25}\\
Y_V &= \big(C_9'^{\text{eff}} - C_9^{\text{eff}}\big)\, f_-(q^2)
      + \frac{2 m_b}{m_B+m_{K_0^*}}\big(C_7^{\text{eff}} - C_7'^{\text{eff}}\big)\,
      \frac{m_B^2 - m_{K_0^*}^2}{q^2}\, f_T(q^2), \label{eq:26}\\
X_A &= \big(C_{10}' - C_{10}\big)\, f_+(q^2),\label{eq:27}\\
Y_A &= \big(C_{10}' - C_{10}\big)\, f_-(q^2).\label{eq:28}
\end{align}

In the SM scenario one has $C_9^{\text{eff}}$, $C_{10}$,
$C_7^{\text{eff}}$ and vanishing chirality-flipped coefficients
$C_{7}'^{\text{eff}}=C_{9}'^{\text{eff}}=C_{10}'=0$. In contrast, in the NP scenario considered here we use
$C_9^{\text{eff}} \to C_9^{\text{eff,tot}}$, \quad $C_9'^{\text{eff}}\to C_9'^{\text{eff,tot}}$,
\quad $C_{10}\to C_{10}^{\text{tot}}$, \quad $C_{10}'\to C_{10}'^{\text{tot}}$,
\quad while $C_7^{\text{eff}} = C_7^{\text{eff,tot}}$ ( $C_7'^{\text{eff}} =0$), remain the same as defined in the effective Hamiltonian.

\section{Differential decay rate}\label{ddr}

The decay rate and angular observables are obtained from the squared matrix element,
summed over final-state spins. It is
convenient to express the result in terms of dimensionless kinematic variables,
\begin{equation}  
\hat s = \frac{q^2}{m_B^2}, \quad \hat m_\ell = \frac{m_\ell}{m_B}, \quad \hat m_{K_0^*} = \frac{m_{K_0^*}}{m_B},
\end{equation}

and the Källén function
\begin{equation}
\lambda(\hat s) = 1 + \hat m_{K_0^*}^4 + \hat s^2 - 2\hat s - 2 \hat m_{K_0^*}^2(1+\hat s), 
\end{equation}
and
\begin{equation}
 v = \sqrt{1 - \frac{4\hat m_\ell^2}{\hat s}}.
 \end{equation}

The variable $v$ represents the velocity of the lepton in the dilepton rest frame.

After integrating over the lepton polar angle, the single differential rate for
$B \to K_0^*(1430)\,\ell^+ \ell^-$ can be written as
\begin{align}
\frac{d\Gamma}{d\hat s} &=
\frac{G_F^2 \alpha_{\rm em}^2 |V_{tb}V_{ts}^*|^2}{2048 \pi^5}
m_B^5 \, \sqrt{\lambda(\hat s)}\, v \,
\bigg[ \frac{\lambda(\hat s) v^2}{3} \Big( |X_V|^2 + |X_A|^2 \Big)
+ \frac{\hat m_\ell^2}{\hat s} \Big( \big|X_V (1-\hat m_{K_0^*}^2) + Y_V \hat s\big|^2 + \big|X_A (1-\hat m_{K_0^*}^2) + Y_A \hat s\big|^2 \Big) \bigg].
\end{align}

The first term, proportional to $\lambda(\hat s)v^2/3$, dominates for light leptons
($\ell=e,\mu$) and encodes the contributions from transverse helicity configurations.
The second term, proportional to $\hat m_\ell^2/\hat s$, becomes relevant for heavier
leptons (e.g.\ $\tau$) and probes combinations of the scalar kinematic structures
$Y_V$ and $Y_A$. In particular, this term can be enhanced in scenarios with large
scalar or pseudoscalar contributions, although such operators are absent in the SM
for light leptons.

\section{Numerical results}\label{sec:results_LFU}
\subsection{Differential decay-rate spectra}

\begin{figure}[p]
\centering

\begin{subfigure}[t]{0.9\textwidth}
\centering
\includegraphics[height=0.26\textheight,keepaspectratio]{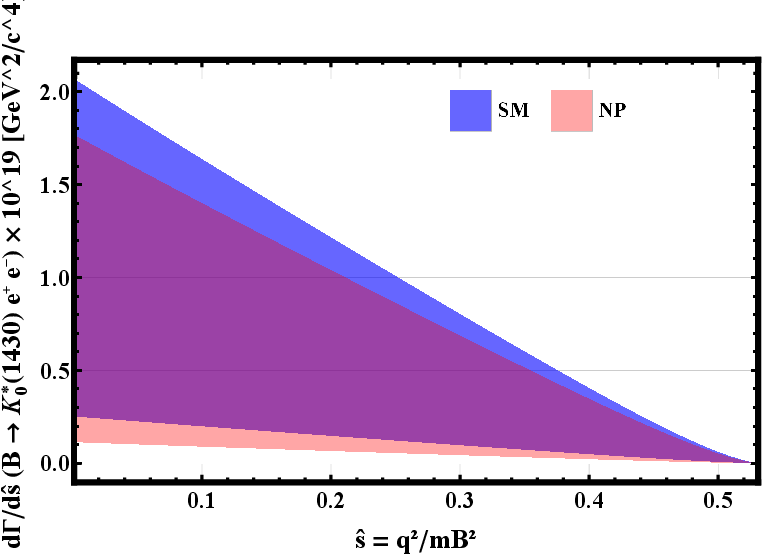}
\caption{$e^+e^-$}
\label{fig:electron}
\end{subfigure}

\vspace{0.6em}

\begin{subfigure}[t]{0.9\textwidth}
\centering
\includegraphics[height=0.26\textheight,keepaspectratio]{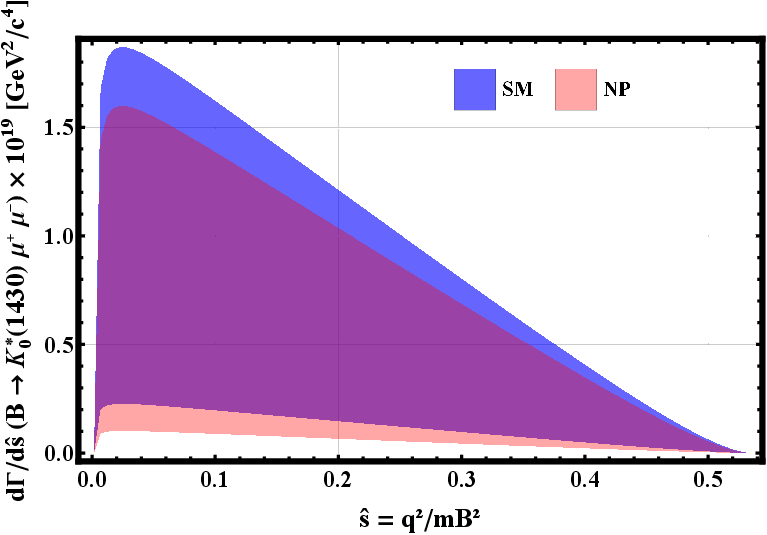}
\caption{$\mu^+\mu^-$}
\label{fig:muon}
\end{subfigure}

\vspace{0.6em}

\begin{subfigure}[t]{0.9\textwidth}
\centering
\includegraphics[height=0.26\textheight,keepaspectratio]{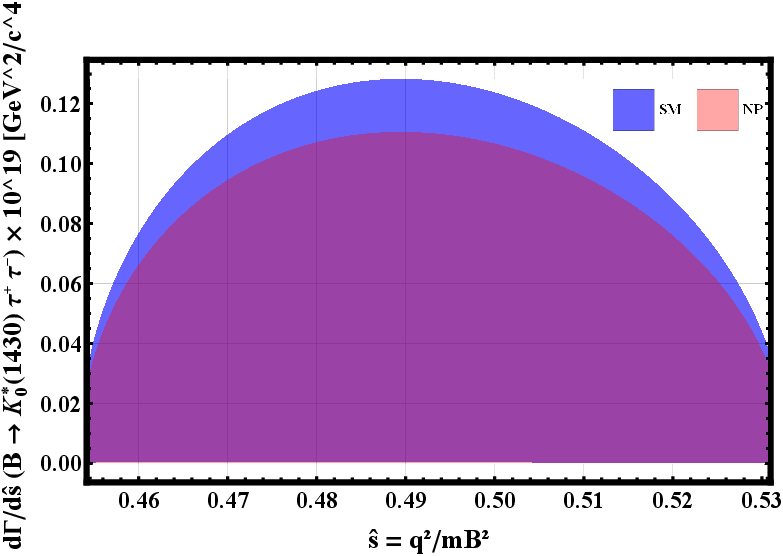}
\caption{$\tau^+\tau^-$}
\label{fig:tau}
\end{subfigure}
\caption{Differential decay rate for $B \to K_0^*(1430)\,\ell^+\ell^-$ in the SM and in the scalar LQ scenario ($\ell=e,\mu,\tau$). For $\tau^+\tau^-$ the spectrum starts at $q^2\ge 4m_\tau^2$, so the low-$q^2$ region is not accessible.}
\label{fig:diff_all}
\end{figure}

Numerically, we use the $B\to K_0^*(1430)$ form factors from Ref.~\cite{Aliev:2007rq} together with
the Wilson coefficients specified in Sec.~\ref{sec:Heff}. The resulting differential spectra
$\mathrm{d}\Gamma/\mathrm{d}\hat s$ with $\hat s \equiv q^2/m_B^2$ are shown in
Figs.~\ref{fig:electron}--\ref{fig:tau} for $\ell=e,\mu,\tau$, comparing the SM with
representative scalar-leptoquark (LQ) benchmarks. In each panel, the blue shaded band represents the
SM prediction and the pink band the LQ scenario; the purple region corresponds to the overlap of the
two semi-transparent bands. The band widths reflect the combined theoretical
uncertainty from the form factors and from varying the Wilson coefficients within the benchmark
ranges of Sec.~\ref{sec:Heff}.  

For $\ell=e,\mu$, the spectra are largest at low $q^2$ and decrease towards the endpoint
$q^2_{\max}=(m_B-m_{K_0^*})^2$, as expected from the shrinking phase space at large dilepton invariant
mass. The enhancement at low $q^2$ is driven by the photon-penguin contribution associated with
$C_7^{\text{eff}}$, which is amplified by the $1/q^2$ factor in the amplitude. At low and
intermediate $q^2$, the semileptonic operators proportional to $C_9^{\text{eff}}$ and $C_{10}$ give
additional contributions and interference terms that shape the spectra. For the scalar-LQ benchmarks
considered here, the predicted rates are typically reduced with respect to the SM over most of the
accessible range (pink band below blue), with the largest relative effect in the region where the
interplay of $C_7^{\text{eff}}$ and $C_9^{\text{eff}}$ is most pronounced.

In the $\tau^+\tau^-$ channel, the spectrum starts only at the physical threshold $q^2\ge 4m_\tau^2$,
so only the high-$q^2$ region is accessible (Fig.~\ref{fig:tau}). In this regime the photon-penguin
term is comparatively less important, and the rate is dominated by the semileptonic contributions
controlled by $C_9^{\text{eff}}$ and $C_{10}$, modulated by the reduced available phase space.
Consequently, the SM and LQ predictions differ mainly through an overall normalization, while the
shape of the spectrum is less strongly modified than in the light-lepton modes. Related analyses of scalar operators in $b\to s\tau^+\tau^-$ transitions, and their
impact on angular observables in $B\to K^{*0}\tau^+\tau^-$, can be found in
Ref.~\cite{Karmakar:2024dml}.

The overall shape of the spectra is largely governed by phase space and the helicity structure of the
$b\to s\ell^+\ell^-$ transition. For $\ell=e,\mu$, the rate is largest at low $q^2$ and decreases
towards the kinematic endpoint $q^2_{\rm max}=(m_B-m_{K_0^*})^2$, reflecting the shrinking phase space
at large dilepton invariant mass. At low and intermediate $q^2$ the decay is mainly driven by the
vector contribution proportional to $C_9^{\text{eff}}$ together with the photon-penguin term governed by
$C_7^{\text{eff}}$, while the axial-vector piece proportional to $C_{10}$ enters through $|C_{10}|^2$
and through interference with the vector amplitude. The scalar-LQ benchmarks considered here modify
these short-distance coefficients, leading predominantly to a suppression of the rate relative to the SM
(red band below blue) over most of the accessible range. The effect is most visible where the
$C_9^{\text{eff}}$--$C_7^{\text{eff}}$ interplay and their interference with $C_{10}$ provide the dominant
contribution, i.e.\ away from the endpoint.

For $\ell=\tau$, the spectrum starts only at threshold $q^2\ge 4m_\tau^2$ and is therefore confined to
the high-$q^2$ region shown in Fig.~\ref{fig:tau}. In this regime the photon-penguin contribution is
less important, and the rate is more strongly shaped by the semileptonic operators governed by
$C_9^{\text{eff}}$ and $C_{10}$ together with the reduced available phase space. As a consequence, the
SM and LQ predictions differ mainly through an overall normalization, while the spectral shape is less
dramatically distorted, consistent with the limited kinematic lever arm in the $\tau^+\tau^-$ channel.

Overall, as is seen, the differential decay rate for the LQ and SM scenarios show  large overlap regions for all leptons.  For $\ell=e,\mu$,  LQ scans some narrow regions out of the SM bands.

\newpage
\subsection{Other observables results}
\label{sec:obs_results}

Having established the differential spectra, we now turn to integrated observables that are directly
relevant for phenomenology and  comparison with experimental data. In particular, we present
numerical predictions for the partially integrated branching fractions  and their ratios at different lepton channels for tests of
lepton-flavor universality (LFU) in $B\to K_0^*(1430)\ell^+\ell^-$, both in the SM and in the
scalar-LQ scenario. These quantities probe complementary combinations of the short-distance Wilson
coefficients and are less sensitive to bin-by-bin fluctuations of the spectrum.

A central issue in $b\to s\ell^+\ell^-$ phenomenology is the presence of long-distance hadronic
effects associated with intermediate charmonium states, which strongly enhance the rate when
$q^2\simeq m_{J/\psi}^2$ or $m_{\psi'}^2$ and can obscure potential short-distance NP contributions.
To reduce this impact, we follow a standard strategy by partition of the kinematically allowed $q^2$
range into three different regions, removing symmetric veto windows around the $J/\psi$ and
$\psi'$ resonances. We then integrate the differential rate over each region to obtain partial
branching fractions, and construct LFU ratios by taking appropriate muon-to-electron (and, where
applicable, tau-to-muon/electron) combinations. In this way, the resulting observables retain high
sensitivity to the LQ-induced shifts in the semileptonic coefficients while minimizing the dominant
charmonium-related uncertainties.

The physical ranges are
\begin{align*}
\textbf{Region I:} \quad & 4m_\ell^2 \le q^2 \le (m_{J/\psi} - 0.02\,\text{GeV})^2, \\
\textbf{Region II:} \quad & (m_{J/\psi} + 0.02\,\text{GeV})^2 \le q^2 \le (m_{\psi'} - 0.02\,\text{GeV})^2, \\
\textbf{Region III:} \quad & (m_{\psi'} + 0.02\,\text{GeV})^2 \le q^2 \le (m_B - m_{K_0^*})^2,
\end{align*}
with $m_{J/\psi}=3.0969$ GeV, $m_{\psi'}=3.6861$ GeV, $m_B=5.279$ GeV, and
$m_{K_0^*}=1.425$ GeV. Region~I corresponds to the low-$q^2$ domain, Region~II to the
intermediate-$q^2$ window between the two charmonium peaks, and Region~III to the
high-$q^2$ endpoint region.

The integrated branching ratios in each region, including uncertainties from the form
factors and variations of the Wilson coefficients within their benchmark ranges, are
collected in Tables~\ref{tab:BR_electron_updated},~\ref{tab:BR_muon_updated} and ~\ref{tab:BR_tautau} for
the $\ell$, $\mu$ and $\tau$ channels, respectively,  where the large relative uncertainty in the $\tau$ channel reflects both the limited
phase space and the enhanced sensitivity to helicity-suppressed contributions.
\begin{table}[htbp]
\centering
\caption{Branching ratios for $B \to K_0^*(1430)\,e^+e^-$ in the SM and in the scalar LQ model.
The quoted ranges reflect combined uncertainties from form factors and Wilson coefficients.}
\label{tab:BR_electron_updated}
\begin{tabular}{lcc}
\hline\hline
Region & SM (Range) & LQ (Range) \\
\hline
Region I (Low $q^2$) & $[1.28-10.57]\times 10^{-8}$ & $[0.57-9.04]\times 10^{-8}$ \\
Region II ($J/\psi$ gap) & $[0.13-1.08]\times 10^{-8}$ & $[0.06-0.93]\times 10^{-8}$ \\
Region III (High $q^2$) & $[0.04-0.31]\times 10^{-9}$ & $[0.02-0.27]\times 10^{-9}$ \\
Total (3 regions) & $[1.43-11.83]\times 10^{-8}$ & $[0.64-10.12]\times 10^{-8}$ \\
\hline\hline
\end{tabular}
\end{table}
\begin{table}[htbp]
\centering
\caption{Branching ratios for $B \to K_0^*(1430)\,\mu^+\mu^-$ in the SM and in the scalar LQ model.
The quoted ranges reflect combined uncertainties from form factors and Wilson coefficients.}
\label{tab:BR_muon_updated}
\begin{tabular}{lcc}
\hline\hline
Region & SM (Range) & LQ (Range) \\
\hline
Region I (Low $q^2$) & $[1.24-10.23]\times 10^{-8}$ & $[0.55-8.75]\times 10^{-8}$ \\
Region II ($J/\psi$ gap) & $[0.13-1.09]\times 10^{-8}$ & $[0.06-0.93]\times 10^{-8}$ \\
Region III (High $q^2$) & $[0.04-0.32]\times 10^{-9}$ & $[0.02-0.27]\times 10^{-9}$ \\
Total (3 regions) & $[1.39-11.50]\times 10^{-8}$ & $[0.62-9.84]\times 10^{-8}$ \\
\hline\hline
\end{tabular}
\end{table}
\FloatBarrier
\begin{table}[htbp]
\centering
\caption{Branching ratios for $B \to K_0^*(1430)\,\tau^+\tau^-$ in the SM and in the scalar LQ model.
Region I is kinematically forbidden for $\tau^+\tau^-$ because $q^2<4m_\tau^2$. The quoted ranges reflect
combined uncertainties from form factors and Wilson coefficients.}
\label{tab:BR_tautau}
\begin{tabular}{lcc}
\hline\hline
Region & SM (Range) & LQ (Range) \\
\hline
Region I (Low $q^2$) & \multicolumn{2}{c}{$q^2<4m_\tau^2$ (forbidden)} \\
Region II ($J/\psi$ gap) & $[0.03-6.08]\times 10^{-10}$ & $[0.01-5.24]\times 10^{-10}$ \\
Region III (High $q^2$) & $[0.02-8.94]\times 10^{-10}$ & $[0.01-7.71]\times 10^{-10}$ \\
Total (physical $q^2$) & $[0.06-18.47]\times 10^{-10}$& $[0.02-15.92]\times 10^{-10}$ \\
\hline\hline
\end{tabular}
\end{table}
\newpage

Although, large intersections between the SM and LQ models are seen, there are some narrow regions for LQ out of the SM predictions considering all the uncertainties.


To test lepton-flavor universality  in $B\to K_0^*(1430)\ell^+\ell^-$, we define
\begin{equation}
R_{K_0^*}\equiv
\frac{\mathcal{B}\!\left(B\to K_0^*(1430)\,\mu^+\mu^-\right)}
{\mathcal{B}\!\left(B\to K_0^*(1430)\,e^+e^-\right)}\,,
\label{eq:RK0star_def}
\end{equation}
where the branching fractions are integrated over the same short-distance $q^2$ regions
(with the $J/\psi$ and $\psi'$ vetoes) as in Tables.~\ref{tab:BR_electron_updated}--\ref{tab:BR_muon_updated}.
Using the corresponding total integrated ranges, we obtain
\begin{align}
R_{K_0^*}^{\rm SM} &\in [0.9673,\,0.9721], \label{eq:RK0star_SM}\\
R_{K_0^*}^{\rm LQ} &\in [0.9659,\,0.9723]. \label{eq:RK0star_LQ}
\end{align}
As expected, $R_{K_0^*}$ is very close to unity in both scenarios. The dominant hadronic
uncertainties largely cancel in the ratio, while the remaining deviation from $1$ is mainly
driven by lepton-mass effects and the slightly different kinematics between the $e$ and $\mu$
modes (together with small residual differences induced by the chosen short-distance windows).
For the scalar-LQ benchmarks considered here, the predicted shift in $R_{K_0^*}$ is modest,
indicating no sizeable LFU violation in the $\mu/e$ sector for these parameter choices.

\vspace{0.4em}
A complementary ratio involving $\tau$ leptons can be defined only in kinematic regions accessible
to both $\mu^+\mu^-$ and $\tau^+\tau^-$. We therefore focus on the high-$q^2$ window (Region~III) and define
\begin{equation}
R_{K_0^*}^{\tau\mu}\big|_{\rm III}\equiv
\frac{\mathcal{B}_{\rm III}\!\left(B\to K_0^*(1430)\,\tau^+\tau^-\right)}
{\mathcal{B}_{\rm III}\!\left(B\to K_0^*(1430)\,\mu^+\mu^-\right)}\,,
\label{eq:RK0star_taumu_def}
\end{equation}
where $\mathcal{B}_{\rm III}$ denotes integration over Region~III only. From
Tables.~\ref{tab:BR_muon_updated} and~\ref{tab:BR_tautau}, we find the indicative ranges
\begin{equation}
R_{K_0^*}^{\tau\mu}\big|_{\rm III,\,SM}\in[4.96\times 10^{-3},\,2.41\times 10^{1}],\qquad
R_{K_0^*}^{\tau\mu}\big|_{\rm III,\,LQ}\in[2.16\times 10^{-3},\,2.08\times 10^{1}]\,.
\label{eq:RK0star_taumu_ranges}
\end{equation}
The wide spread is a direct consequence of the much larger relative uncertainty in the
$\tau^+\tau^-$ mode: since $q^2_{\max}=(m_B-m_{K_0^*})^2$ lies only moderately above the threshold
$4m_\tau^2$, the available phase space is compressed near the endpoint, enhancing the sensitivity
to form-factor variations and to helicity-suppressed contributions. As a result, $R_{K_0^*}^{\tau\mu}$
is less constraining in a conservative range scan, but it remains a useful high-$q^2$ LFU probe once
uncertainties are reduced and more targeted benchmark choices are adopted.

\section{Forward-backward asymmetry and lepton polarization}\label{fba}

Angular observables provide additional, and often more reliable, probes of the underlying
short-distance dynamics. In particular, the forward-backward asymmetry of the lepton and
the lepton polarization asymmetries are sensitive to the chiral and scalar structure of
the effective Hamiltonian and can help disentangle different NP scenarios.

The double differential decay width in the dilepton invariant mass $q^2$ and the angle
$\theta$ between the momentum of the negatively charged lepton $\ell^-$ and the $K_0^*$
in the dilepton rest frame can be written as
\[
\frac{d^2\Gamma}{dq^2\,d\cos\theta}
= \frac{1}{512\pi^3 m_B^3}\,
\lambda^{1/2}(m_B^2,m_{K_0^*}^2,q^2)\,
\beta_\ell\;
\sum_{\rm spins}|\mathcal M|^2,
\]
with the usual kinematic functions
\[
\beta_\ell = \sqrt{1-\frac{4m_\ell^2}{q^2}},\qquad
L\equiv\frac{\beta_\ell\sqrt{\lambda}}{4}\,.
\]

After carrying out the spin sums and expressing the result in terms of $X_{V,A}$ and
$Y_{V,A}$, the angular distribution of the $B \to K_0^* \ell^+ \ell^-$ decay can be cast
in the compact form
\begin{equation}
\frac{d^2 \Gamma}{dq^2 \, d\cos\theta} = \mathcal{P}(q^2) \left[ \mathcal{C}_0(q^2) + \mathcal{C}_2(q^2)\cos^2\theta \right],
\end{equation}
where the common prefactor is
\begin{equation}
\mathcal{P}(q^2)=\frac{1}{512\pi^3 m_B^3}\,\lambda^{1/2}(m_B^2,m_{K_0^*}^2,q^2)\,\beta_\ell
\cdot 4|N|^2,
\end{equation}
and $N=\dfrac{G_F\alpha_{\rm em}V_{tb}V_{ts}^*}{2\sqrt2\pi}$.

Defining the auxiliary combinations
\begin{align}
V\!\cdot\!V &= X_V^{2}\,P^{2} + 2\,X_V Y_V\,\Delta + Y_V^{2}\,q^{2}, \\
A\!\cdot\!A &= X_A^{2}\,P^{2} + 2\,X_A Y_A\,\Delta + Y_A^{2}\,q^{2}, \\
P^{2} &= 2\!\left(m_B^{2}+m_{K_0^*}^{2}\right) - q^{2},
\end{align}
one finds
\begin{align}
\mathcal{C}_{0}(q^{2})
&= 2\Big[\big(X_V K + Y_V Q\big)^{2} + \big(X_A K + Y_A Q\big)^{2}\Big]
- \frac{1}{2}\,q^{2}\big(V\!\cdot\!V + A\!\cdot\!A\big), \\[4pt]
\mathcal{C}_{2}(q^{2})
&= -2\,L^{2}\,\big(X_V^{2}+X_A^{2}\big),
\end{align}
where $K$ and $Q$ are simple kinematic functions of the masses and $q^2$ (their explicit
expressions can be found from the full decomposition of the amplitude but are not
needed here).

The forward-backward asymmetry (FBA) is defined in the standard way as
\begin{equation}
A_{FB}(q^{2}) =
\frac{
\displaystyle \int_{0}^{1} d\cos\theta\, \frac{d^{2}\Gamma(q^{2},\cos\theta)}{dq^{2}\, d\cos\theta}
- \int_{-1}^{0} d\cos\theta\, \frac{d^{2}\Gamma(q^{2},\cos\theta)}{dq^{2}\, d\cos\theta}
}{
\displaystyle \int_{0}^{1} d\cos\theta\, \frac{d^{2}\Gamma(q^{2},\cos\theta)}{dq^{2}\, d\cos\theta}
+ \int_{-1}^{0} d\cos\theta\, \frac{d^{2}\Gamma(q^{2},\cos\theta)}{dq^{2}\, d\cos\theta}
}.
\end{equation}

For a scalar final state such as $K_0^*(1430)$, and in the absence of scalar-type
lepton couplings, the angular dependence reduces to an even function of $\cos\theta$.
It follows that the numerator of $A_{FB}(q^2)$ vanishes identically in the SM, and
$A_{FB}(q^2)=0$ for all $q^2$~\cite{Aslam:2009cv}. This makes the FBA in
$B \to K_0^*(1430)\ell^+\ell^-$ a particularly clean null test of the SM: any non-zero
measurement would point directly to NP with scalar or pseudoscalar couplings.

\vspace{0.2cm}

In addition to the FBA, lepton polarization asymmetries provide complementary
information on the chiral structure of the underlying short-distance dynamics.
For a given spin direction $\hat{\mathbf{i}}$ ($i=L,T,N$), we define
\begin{equation}
P_i(q^2)=
\frac{\displaystyle \frac{d\Gamma}{dq^2}(\hat{\mathbf{s}}_- = +\hat{\mathbf{i}})-
\frac{d\Gamma}{dq^2}(\hat{\mathbf{s}}_- = -\hat{\mathbf{i}})}
{\displaystyle \frac{d\Gamma}{dq^2}(\hat{\mathbf{s}}_- = +\hat{\mathbf{i}})+
\frac{d\Gamma}{dq^2}(\hat{\mathbf{s}}_- = -\hat{\mathbf{i}})}\,,
\end{equation}
where $L,T,N$ denote the longitudinal, transverse, and normal polarizations of $\ell^-$,
respectively. We work in the dilepton rest frame, choose the $z$-axis along the
$K_0^*$ momentum, and define the $x$--$z$ plane as the decay plane spanned by
$\vec p_{K_0^*}$ and $\vec p_{\ell^-}$.
The spin unit vectors in the $\ell^-$ rest frame are taken as
\begin{align}
\hat{\mathbf{s}}_L &= \frac{\vec p_{\ell^-}}{|\vec p_{\ell^-}|}\,, \\
\hat{\mathbf{s}}_N &= \frac{\vec p_{K_0^*} \times \vec p_{\ell^-}}
{|\vec p_{K_0^*} \times \vec p_{\ell^-}|}\,, \\
\hat{\mathbf{s}}_T &= \hat{\mathbf{s}}_N \times \hat{\mathbf{s}}_L\,.
\end{align}

Using the amplitude in Eq.~(\ref{eq:24}) and the invariant functions $X_V,Y_V,X_A,Y_A$
defined in Eqs.~(\ref{eq:25})--(\ref{eq:28}), the longitudinal polarization of $\ell^-$ in the
\emph{integrated} dilepton spectrum can be written in the compact form
\begin{equation}
P_L(\hat s)=
\frac{4\,v\,\Re\!\left[\,X_V(\hat s)\,X_A^{*}(\hat s)\right]}
{2\,v^{2}\,|X_V(\hat s)|^{2}+\left(3-v^{2}\right)\,|X_A(\hat s)|^{2}}\,.
\label{eq:PL_compact}
\end{equation}

For the decay $B\to K_0^*\ell^+\ell^-$ (spinless hadronic final state), the normal and
transverse polarizations are canceled after angular integration, so that
\begin{equation}
P_N(\hat s)=0\,,\qquad P_T(\hat s)=0\,,
\label{eq:PTPN_zero}
\end{equation}
and $P_L$ is the only non-trivial single-lepton polarization observable in the
integrated dilepton spectrum.

\begin{figure}[htbp]
\centering

\begin{subfigure}{0.75\textwidth}
\centering
\includegraphics[width=\linewidth,height=6.2cm]{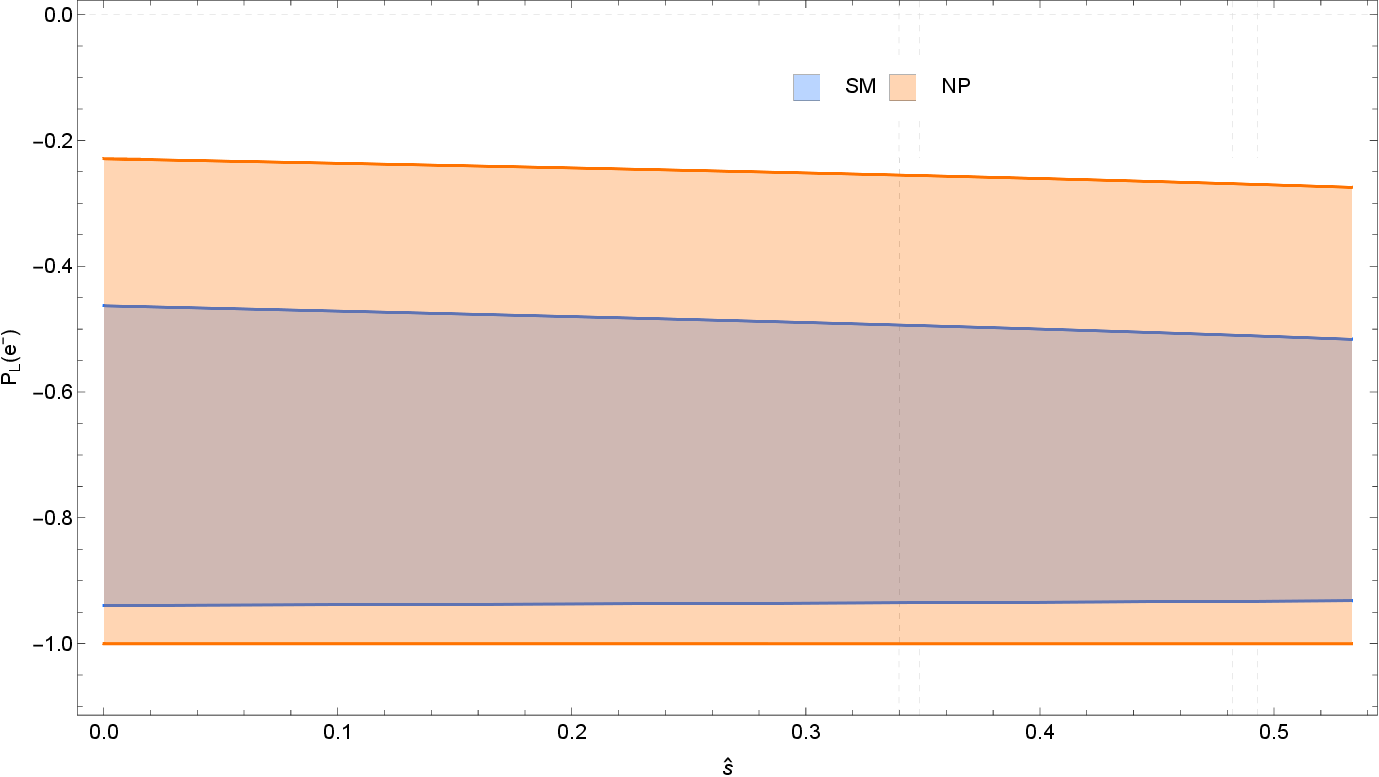}
\caption{Longitudinal lepton polarization asymmetry $P_L$ for $B\to K_0^*(1430)\,e^+e^-$ in the SM and the scalar leptoquark scenario.}
\label{fig:PL_e}
\end{subfigure}

\vspace{0.35cm}

\begin{subfigure}{0.75\textwidth}
\centering
\includegraphics[width=\linewidth,height=6.2cm]{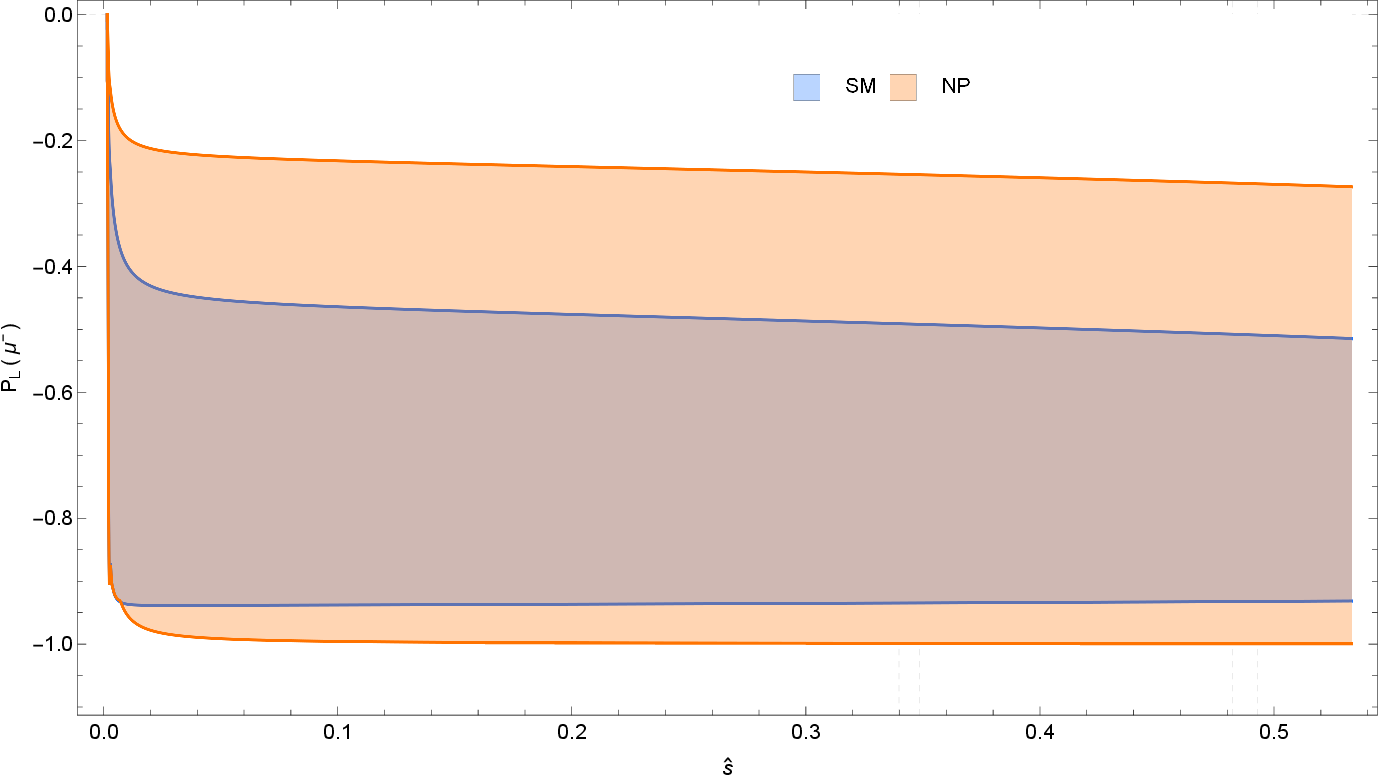}
\caption{Longitudinal lepton polarization asymmetry $P_L$ for $B\to K_0^*(1430)\,\mu^+\mu^-$ in the SM and the scalar leptoquark scenario.}
\label{fig:PL_mu}
\end{subfigure}

\vspace{0.35cm}

\begin{subfigure}{0.75\textwidth}
\centering
\includegraphics[width=\linewidth,height=6.2cm]{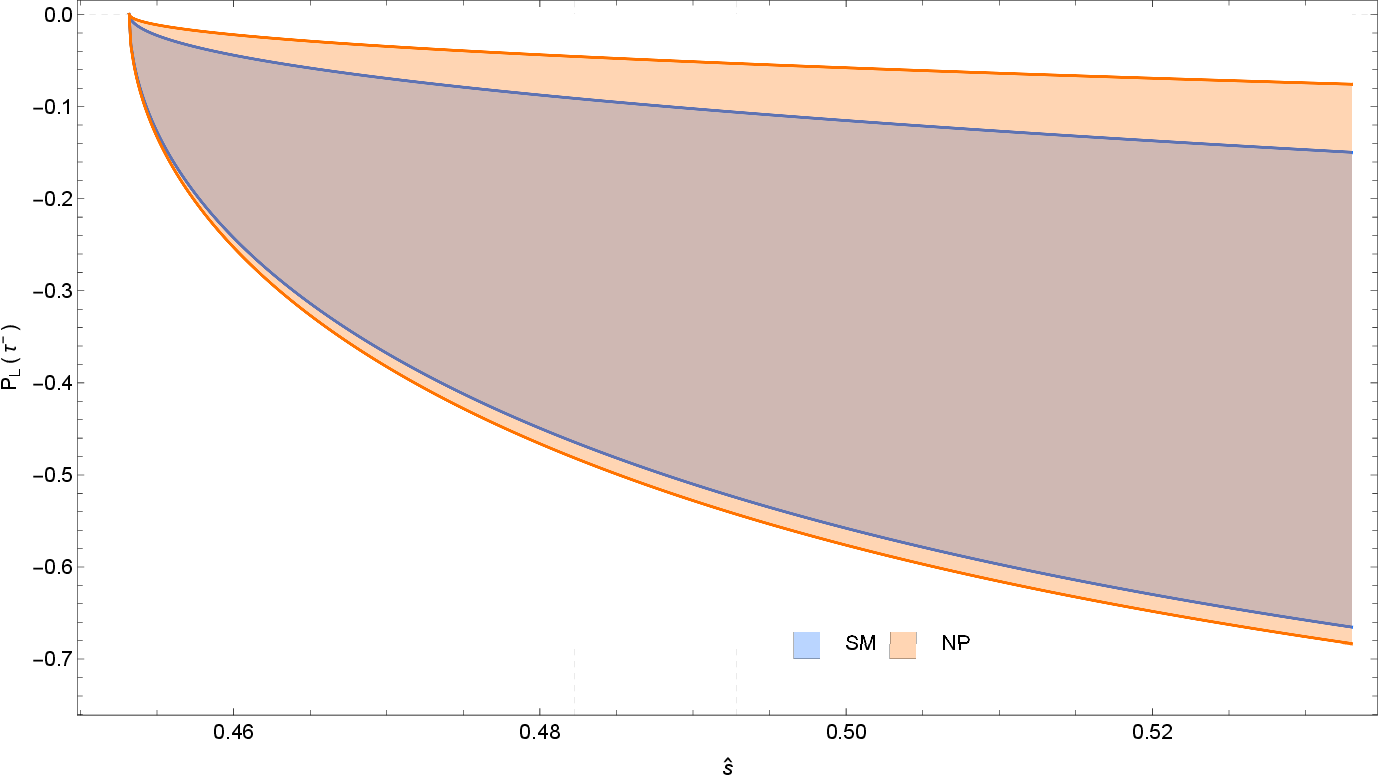}
\caption{Longitudinal lepton polarization asymmetry $P_L$ for $B\to K_0^*(1430)\,\tau^+\tau^-$ in the SM and the scalar leptoquark scenario.}
\label{fig:PL_tau}
\end{subfigure}

\caption{Longitudinal lepton polarization asymmetry $P_L(\hat s)$ in $B\to K_0^*(1430)\,\ell^+\ell^-$ for $\ell=e,\mu,\tau$, comparing the SM prediction with the scalar leptoquark scenario.}
\label{fig:PL_column}
\end{figure}

The longitudinal lepton polarization asymmetry $P_L(\hat s)$ for the decays
$B \to K_0^*(1430)\,\ell^+\ell^-$ is shown in Fig.~\ref{fig:PL_column} as a function of the
dilepton invariant mass squared $q^2$ for $\ell=e,\mu,\tau$.
The individual panels correspond to the electron, muon, and tau channels in
Figs.~\ref{fig:PL_e}, \ref{fig:PL_mu}, and \ref{fig:PL_tau}, respectively.
In each case, the SM prediction is compared with the scalar leptoquark scenario,
illustrating the sensitivity of $P_L$ to the underlying short-distance dynamics.

Since $P_L(\hat s)$ is defined as a ratio of polarized decay rates, it is bounded by
\(-1 \le P_L(\hat s) \le 1\). Negative values correspond to a predominantly left–handed
$\ell^-$, which is the expected pattern in the SM because the weak interaction is
chiral \cite{Chen:2007na}.

Figure~\ref{fig:PL_column} shows $P_L(\hat s)$ for $B\to K_0^*(1430)\,\ell^+\ell^-$ with
$\ell=e,\mu,\tau$, comparing the SM prediction (blue band) to the scalar leptoquark
scenario (orange band). For the light-lepton modes,
$B\to K_0^*(1430)\,e^+e^-$ and $B\to K_0^*(1430)\,\mu^+\mu^-$
(Figs.~\ref{fig:PL_e} and~\ref{fig:PL_mu}), the SM result stays  close to
$P_L(\hat s)\simeq -1$ across almost the entire kinematic range, indicating that the
final-state lepton is produced nearly purely left–handed.
In the leptoquark case the polarization remains negative, but the magnitude is reduced,
so that typically, considering the averages,  $|P_L^{\rm NP}(\hat s)|<|P_L^{\rm SM}(\hat s)|$ throughout the accessible
$\hat s$ region. Physically, this reflects the presence of additional contributions in the
NP scenario that populate the opposite-helicity component and therefore dilute the net
longitudinal polarization \cite{Chen:2007na}.

The situation changes noticeably for the $\tau$ channel in Fig.~\ref{fig:PL_tau}.
Because $m_\tau$ is large, helicity suppression is much less effective, and the SM prediction
develops a clear $\hat s$ dependence, with $P_L(\hat s)$ departing significantly from $-1$
towards the upper end of phase space.
In the scalar leptoquark scenario this effect becomes even more pronounced: scalar-type
contributions scale with the lepton mass and are therefore most important for $\ell=\tau$,
leading to a smaller $|P_L(\hat s)|$.  As a result, the longitudinal polarization in
$B\to K_0^*(1430)\,\tau^+\tau^-$ offers a particularly sensitive handle on scalar leptoquark
effects in rare $B$ decays \cite{Chen:2007na}.

In summary, $P_L(\hat s)$ is nearly saturated at $-1$ and only weakly depends on $\hat s$
for $\ell=e,\mu$, while for $\ell=\tau$ both the kinematic variation and the size of possible
deviations from the SM are  enhanced,  especially at large $\hat s$.
\newpage
\section{Conclusions}\label{conclu}

In this work, we studied the rare decay
$B\to K_0^*(1430)\,\ell^+\ell^-$ with $\ell=e,\mu,\tau$ as a sensitive probe
of new physics in $b\to s\ell^+\ell^-$ transitions, focusing on scalar
leptoquark scenarios that can accommodate current flavor anomalies and the
muon anomalous magnetic moment. On the SM side, we expressed
the semileptonic amplitude in terms of the usual effective coefficients
and concentrated on short-distance $q^2$ windows. By imposing explicit
vetoes around the $J/\psi$ and $\psi'$ resonances, we strongly suppressed
long-distance charmonium effects and kept theoretical uncertainties under
control. Using QCD sum-rule form factors for the $B\to K_0^*(1430)$
transition, we derived analytic expressions and presented numerical
predictions for the differential decay rate, partially integrated
branching fractions, LFU ratios, and lepton polarization observables in
these clean kinematic regions.

We then analyzed the same observables in a scalar-leptoquark framework in
which tree level semileptonic interactions induce shifts in $C_9$,
$C_{10}$ and their chirality-flipped counterparts. In this setup we find
sizeable, but still experimentally realistic, departures from SM
expectations. For the light lepton modes, the branching fractions
integrated over the three short-distance regions fall in the ranges
$\mathcal{B}(B\to K_0^*(1430)e^+e^-)_{\rm SM}\simeq
(1.4$--$11.8)\times10^{-8}$ and
$\mathcal{B}(B\to K_0^*(1430)\mu^+\mu^-)_{\rm SM}\simeq
(1.4$--$11.5)\times10^{-8}$,
while the scalar leptoquark benchmarks typically lower them to about
$(0.6$--$10.1)\times10^{-8}$ and $(0.6$--$9.8)\times10^{-8}$,
respectively. In other words, within the quoted uncertainties, the new
physics benchmarks can suppress the average  rates by up to
$\mathcal{O}(20\%)$ across the short-distance regions. The $\tau$ mode is
much rarer,
$\mathcal{B}(B\to K_0^*(1430)\tau^+\tau^-)\sim(0.6$--$1.8)\times10^{-9}$ in
the SM and $(0.2$--$1.6)\times10^{-9}$ in the leptoquark scenario, but it
remains particularly informative because helicity-suppressed effects are
less severe and the decay is therefore more sensitive to the chiral
structure of possible new interactions.

Lepton flavor universality tests constructed from the electron and muon
channels, and especially
\[
R_{K_0^*}\equiv
\frac{\mathcal{B}(B\to K_0^*(1430)\mu^+\mu^-)}
{\mathcal{B}(B\to K_0^*(1430)e^+e^-)}\,,
\]
stay extremely close to unity in both scenarios:
$R_{K_0^*}^{\rm SM}\in[0.9673,0.9721]$ and
$R_{K_0^*}^{\rm LQ}\in[0.9659,0.9723]$.
This behaviour is expected: hadronic uncertainties largely cancel in the
ratio and lepton mass effects are tiny in the $\mu/e$ sector. As a
result, the benchmark leptoquark couplings considered here do not predict
large LFU violation between electrons and muons in this channel. Ratios
involving $\tau$ leptons, such as $R_{K_0^*}^{\tau\mu}$ in the high $q^2$
region, span a wider range because the available phase space is strongly
compressed near threshold. Nevertheless, with improved form factor input
and more targeted benchmark choices, these observables could become
useful probes of non-universality at high $q^2$.

Angular and polarization observables provide an especially clean handle on
short-distance dynamics in this decay. Since the hadronic final state is
a scalar and the SM does not contain scalar lepton couplings, the
forward-backward asymmetry $A_{FB}(q^2)$ vanishes identically for all
$q^2$. Therefore, any nonzero measurement of $A_{FB}$ in
$B\to K_0^*(1430)\ell^+\ell^-$ would be a direct indication of
non-standard scalar or pseudoscalar interactions. The longitudinal lepton
polarization $P_L(\hat s)$ is also very distinctive: for $\ell=e,\mu$ it is
nearly maximally negative and only weakly dependent on $\hat s$ in the SM,
whereas scalar leptoquarks reduce $|P_L|$ by enhancing the opposite
helicity component, producing a characteristic shift that is largely
insensitive to hadronic input. For $\ell=\tau$, $P_L(\hat s)$ shows a much
stronger $\hat s$ dependence and an even clearer separation between SM and
leptoquark predictions, making the $\tau^+\tau^-$ mode a particularly
sensitive probe of scalar new physics in the high--$q^2$ region.

To summarize, $B\to K_0^*(1430)\ell^+\ell^-$ provides a theoretically clean
and phenomenologically rich complement to the better studied
$B\to K^{(*)}\ell^+\ell^-$ modes. In the short-distance windows that will
be accessible at Belle~II and the upgraded LHCb, improved measurements of
differential rates, integrated branching fractions, LFU ratios, and
especially longitudinal lepton polarization can either reveal scalar
leptoquark effects or place strong constraints on them, as well as on
other extensions of the SM that modify the chiral or scalar structure of
$b\to s\ell^+\ell^-$. On the theory side, more precise nonperturbative
determinations of the $B\to K_0^*(1430)$ form factors and a refined
treatment of residual long-distance effects will further sharpen these
predictions and strengthen the discovery potential of this channel in the
high luminosity era.

\section*{ACKNOWLEDGMENTS}

K.~Azizi thanks the Iran National Science Foundation (INSF) for partial financial support provided under the Elites Grant No.~40405095.

\end{document}